\documentclass[3p,times,procedia]{elsarticle}
\usepackage{natbib,hyperref}
\usepackage{epsfig}
\usepackage{amsfonts}
\usepackage{amsmath}
\usepackage{nupha_ecrc}
\newcommand{\rmi}[1]{{\mbox{\scriptsize #1}}}
\newcommand{\rmii}[1]{{\mbox{\tiny\rm{#1}}}}

\newcommand{\nB}{n_\rmii{B}}
\def\gsi{\raise0.3ex\hbox{$>$\kern-0.75em\raise-1.1ex\hbox{$\sim$}}}
\newcommand{\gsim}{\mathop{\gsi}}

\volume{00}

\firstpage{1}

\journalname{Nuclear Physics A}

\runauth{O. Kaczmarek}

\jid{nupha}

\jnltitlelogo{Nuclear Physics A}

\usepackage{amssymb}
\begin{document}

\begin{frontmatter}



%% Instructions from Editor: Please use the following \dochead only in the preprint version (e-print arXiv etc.); 
  %% use empty \dochead{} when submitting to Nuclear Physics A!
\dochead{XXVIth International Conference on Ultrarelativistic Nucleus-Nucleus Collisions\\ (Quark Matter 2017)}
%\dochead{ \ \\ \ }
%% e.g. \dochead{17th International Conference on Dynamical Processes in Excited States of Solids}

\title{Lattice QCD results on soft and hard probes\\of strongly interacting matter}


\author{Olaf Kaczmarek}

\address{Key Laboratory of Quark \& Lepton Physics (MOE) and Institute of
  Partical Physics,\\ Central China Normal University, Wuhan 430079, China;\\
Fakult\"at f\"ur Physik, Universit\"at Bielefeld, D-33615 Bielefeld, Germany}

\begin{abstract}
  We present recent results from lattice QCD relevant for the study of strongly interacting matter as it is produced in heavy ion collision experiments. The equation of state at non-vanishing density from a Taylor expansion up to $6^{th}$ order will be discussed for a strangeness neutral system and using the expansion coefficients of the series limits on the critical point are estimated.
  Chemical freeze-out temperatures from the STAR and ALICE Collaborations will be compared to lines of constant physics calculated from the Taylor expansion of QCD bulk thermodynamic quantities.
We show that qualitative features of the $\sqrt{s_{NN}}$ dependence of skewness and kurtosis ratios of net proton-number fluctuations measured by the STAR Collaboration can be understood from QCD results for cumulants of conserved baryon-number fluctuations. As an example for recent progress towards the determination of spectral and transport properties of the QGP from lattice QCD, we will present constraints on the thermal photon rate determined from a spectral reconstruction of continuum extrapolated lattice correlation functions in combination with input from most recent perturbative calculations. 
\end{abstract}


\end{frontmatter}

\section{Introduction}
The understanding of strongly interacting matter at high temperatures and densities is a major goal of many theoretical studies as well as investigations using heavy ion collision experiments. There has been considerable progress in both directions. The nature and location of the cross-over transition \cite{Bazavov:2011nk,Aoki:2006br,Aoki:2009sc,Borsanyi:2010bp} and the equation of state for vanishing baryon density \cite{Borsanyi:2013bia,Bazavov:2014pvz} are known from continuum extrapolated lattice QCD calculations and extending these studies into the region of the phase diagram at non-vanishing density is constantly progressing (see \cite{Ding:2015ona,Schmidt:2017bjt} for current overviews).\\
Although the hadron resonance gas (HRG) model is able to describe thermodynamic quantities at low temperatures reasonably well, at least when additional yet unobserved resonances are taken into account \cite{Bazavov:2014xya,Alba:2017mqu}, systematic deviations become large already at the lower end of the cross-over region as observed in 
higher order net baryon-number cumulants \cite{Bazavov:2017dus}.
\begin{figure*}[thbp]
%        \centering
        \hspace*{-0.8cm}
    \includegraphics[height=0.25\textwidth]{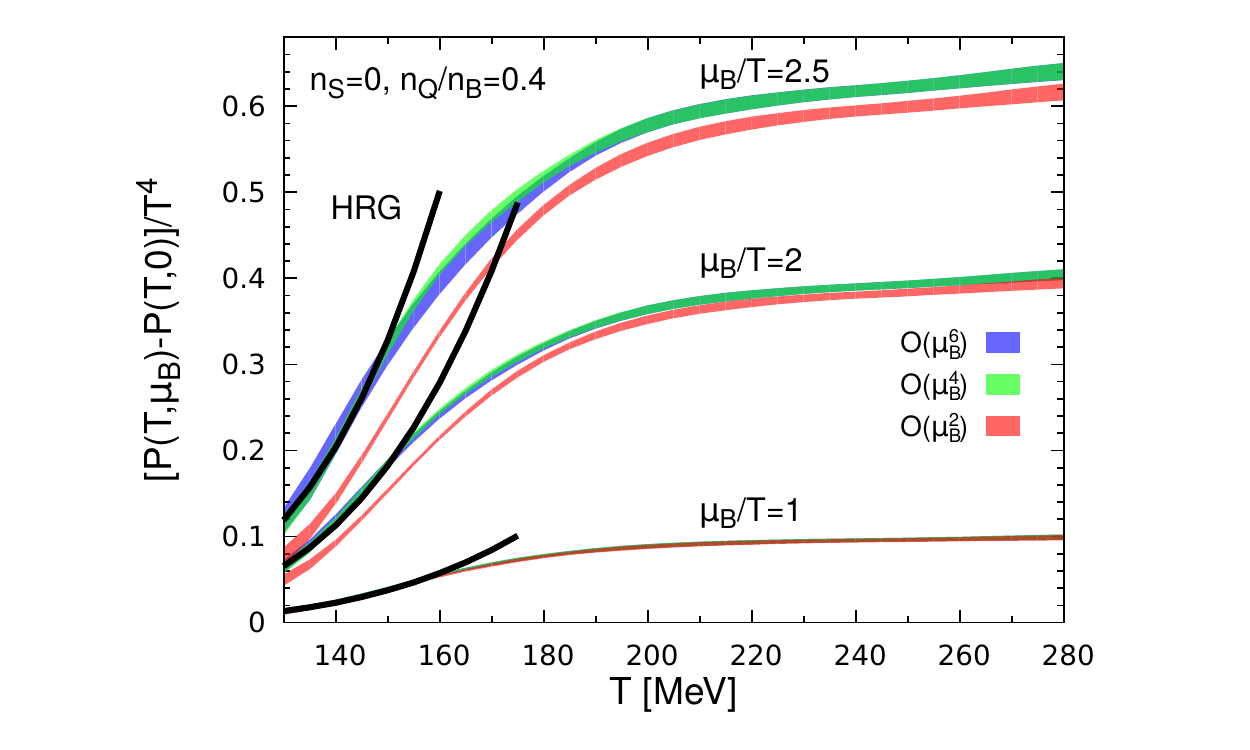}
        \hspace*{-1.6cm}
        \includegraphics[height=0.25\textwidth]{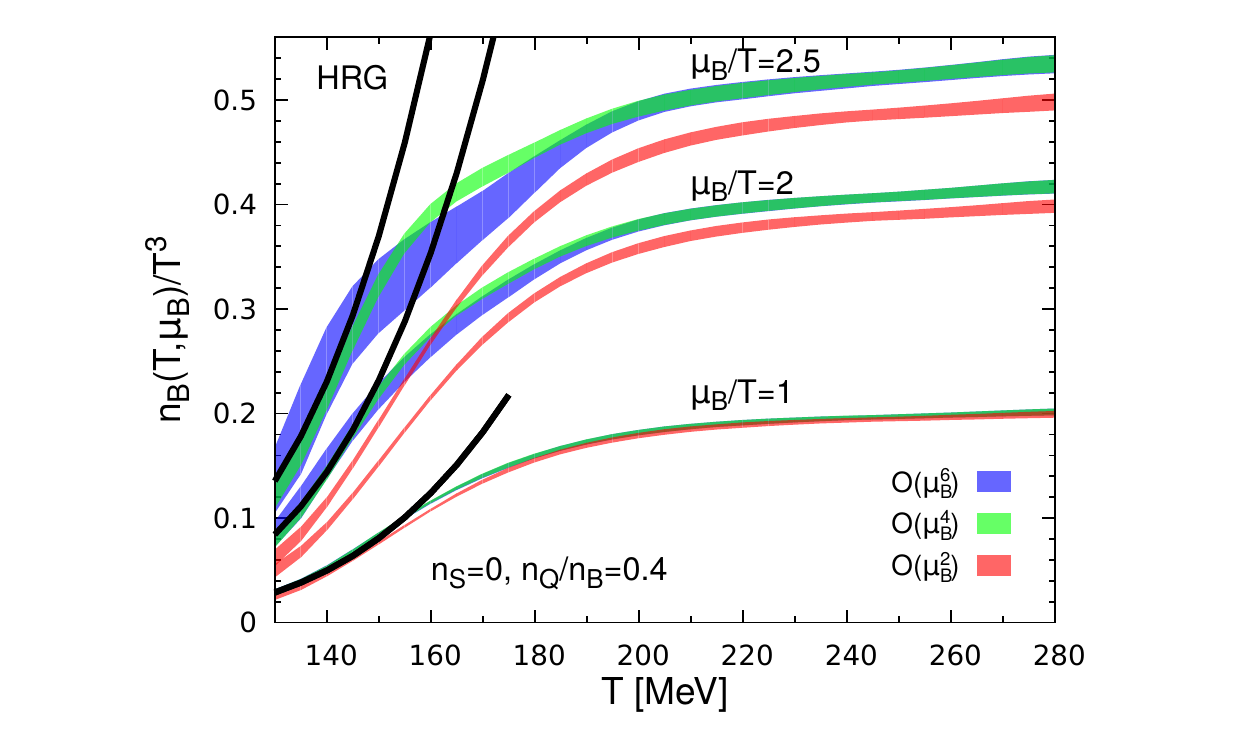}
        \hspace*{-0.9cm}
        \includegraphics[height=0.25\textwidth]{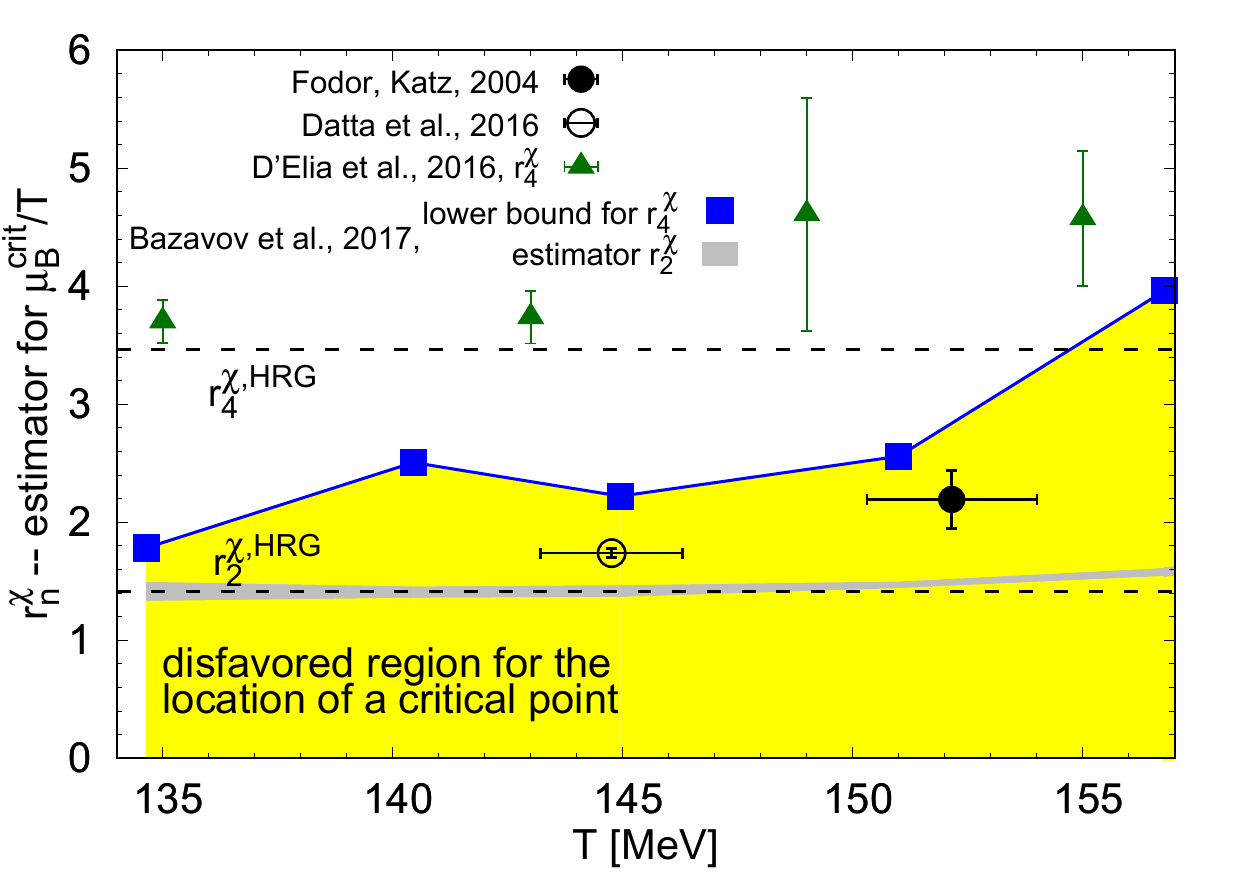}
        \caption{
The $\mu_B$-dependent part of the pressure (left) and baryon-number density (middle)
for a strangeness neutral system with $n_B/n_Q=0.4$.
Estimators for the radius of convergence of the Taylor series for net baryon-number fluctuations, $\chi_2^B(T,\mu_B)$, (right).
Our estimates for $\mu_s=\mu_q=0$ from $N_\tau=8$ from \cite{Bazavov:2017dus} are
compared with results obtained with imaginary chemical potential \cite{DElia:2016jqh},
reweighting technique \cite{Fodor:2004nz} and Taylor expansions \cite{Datta:2016ukp} using standard
staggered fermions, all rescaled using $T_c=154~MeV$.
All figures are taken from \cite{Bazavov:2017dus}.}
        \label{fig_eos}
\end{figure*}
This shows that a Skellam distribution is only valid
at most up to temperatures around 155~MeV consistent with recent measurements of second order cumulants of net proton-number fluctuations by the ALICE Collaboration \cite{Rustamov:2017lio}. Therefore in the regime close to the freeze-out, where conserved charge fluctuations may provide insight into the existence and location of a possible critical point in the QCD phase diagram, it is important to use QCD rather than HRG model calculations.
Lattice QCD results on higher order cumulants of conserved charges provide important insights into the physics at freeze-out, e.g. cumulants of net baryon-number fluctuations can qualitatively describe the features observed from net proton-number fluctuations measured by the STAR Collaboration in the beam energy scan \cite{BNLBI_SKEW} (see section \ref{sec_skew}).\\
Recent progress in the determination of spectral and transport properties of the QGP from continuum extrapolated lattice  correlation functions in combination with recent progress in perturbative calculations that allow to provide constraints on the spectral function at perturbatively high frequencies, allowed to calculate continuum estimates for the thermal dilepton \cite{Ding:2016hua} and photon \cite{Ghiglieri:2016tvj} rates (see section \ref{sec_photon}) as well as transport coefficients like the electrical conductivity \cite{Ding:2016hua} and heavy quark momentum diffusion coefficient \cite{Francis2015II}.
Although still limited to the quenched approximation, these studies, using continuum correlation functions
in combination with phenomenological and perturbatively inspired Ans\"atze,
demonstrate the potential
for a more controlled way to determine spectral and transport properties.
Although not yet in the continuum and with still unphysical quark masses, the effect of dynamical quarks on the electrical conductivity were studied in \cite{Brandt2013,Brandt:2015aqk,Amato2013}.
For a comparison to other calculations see \cite{Greif:2016skc}.

\section{Equation of state and limits on the critical point}
\label{sec_eos}
\begin{figure*}[thbp]
        \centering
        \includegraphics[height=0.36\textwidth]{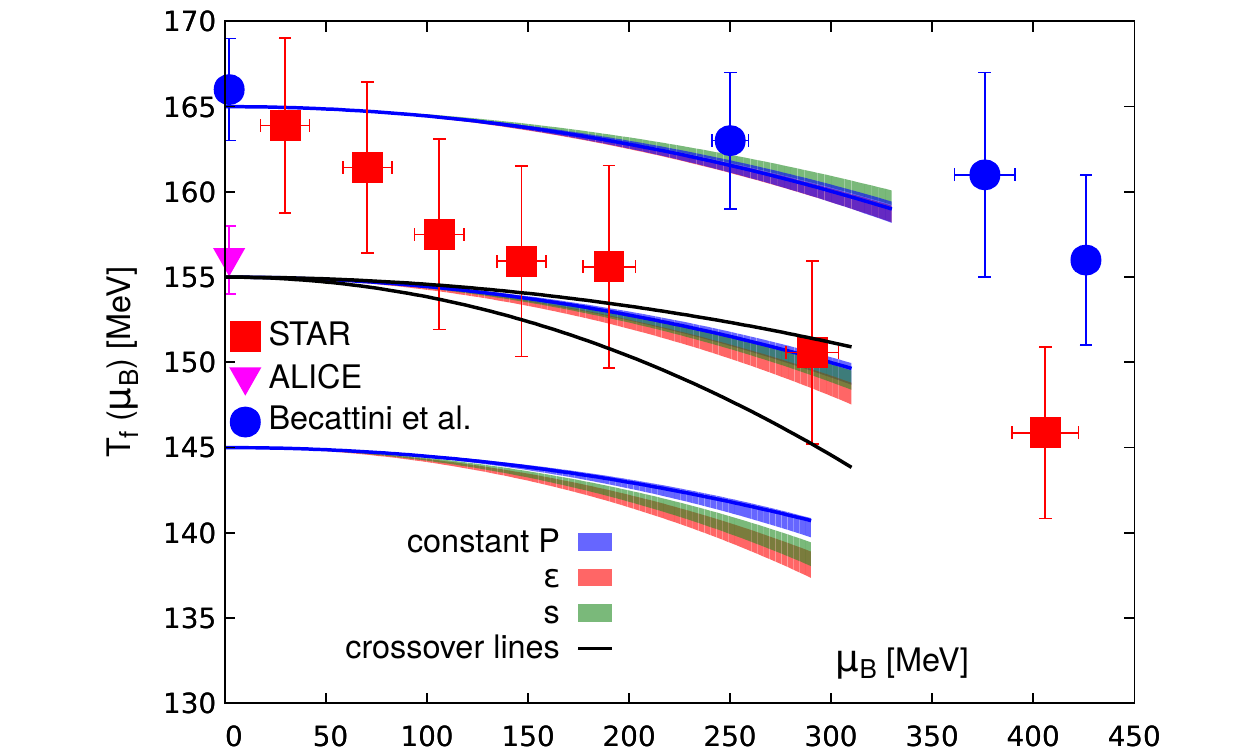}
        \caption{
Comparison of freeze-out temperatures determined by the STAR Collaboration in
the BES at RHIC \cite{Adamczyk:2017iwn} and the ALICE Collaboration at LHC \cite{Floris:2014pta} to lines of
constant pressure, energy density and entropy density in (2+1)-flavor QCD \cite{Bazavov:2017dus}
and hadronization temperatures from a hadronization model calculation \cite{Becattini:2016xct}.
The crossover lines represent the current estimates of the $\mu_B$-dependence
of the QCD crossover transition from \cite{Kaczmarek:2011zz,Endrodi:2011gv,Bonati:2015bha,Bellwied:2015rza,Cea:2015cya}. Figure taken from \cite{Bazavov:2017dus}.
}
        \label{fig_lcp}
\end{figure*}
The equation of state (EoS) of strong-interaction matter, i.e. the temperature and density dependence of bulk thermodynamic quantities, provides the characterization of equilibrium properties and is an important input for hydrodynamic modeling of the evolution of the matter produced in heavy ion collisions. Using continuum extrapolated lattice QCD results with physical light and strange quark masses
consistent results on the EoS of QCD have recently been obtained by two groups 
for vanishing chemical potentials \cite{Borsanyi:2013bia,Bazavov:2014pvz}.
While appropriate for LHC energies, the conditions at the RHIC beam energy scan (BES) require the knowledge of the EoS at non-vanishing baryon ($\mu_B$), strangeness ($\mu_S$) and electric charge ($\mu_Q$) chemical potentials. The strangeness neutrality ($n_S=0$) and fixed net electric-charge to net baryon-number ratio ($n_Q/n_B\simeq 0.4$) in heavy ion collisions relate the different chemical potentials. For small values of the chemical potentials a Taylor expansion allows to extend calculations of thermodynamic quantities for this constraint condition \cite{Bazavov:2017dus}.
In Fig.~\ref{fig_eos} the $\mu_B$ dependent contribution to the pressure (left) and baryon-number density (middle) for the strangeness neutral system are shown up to ${\cal O}(\mu_B^6)$ and three values of $\mu_B$. Comparable results are reported in \cite{Gunther:2016vcp,Guenther17}. For both quantities a good convergence of the Taylor series for $\mu_B\leq 2 T$ even for small temperatures can be observed.
A parametrizations of basic thermodynamic quantities that can readily be incorporated in hydrodynamic simulation codes can be found in \cite{Bazavov:2017dus}.\\ 
Ratios of subsequent expansion coefficients of the Taylor series can be used to estimate the radius of convergence and therefore constraints on the location of a possible critical point in the QCD phase diagram. Recent results on such estimators for the Taylor series up to 6th-order of net baryon-number fluctuations
\cite{Bazavov:2017dus,DElia:2016jqh,Fodor:2004nz,Datta:2016ukp}
are shown in Fig.~\ref{fig_eos} (right). Although a determination of the critical point with these low-order coefficients is not possible,
and even an infinite radius of convergence can not be ruled out,
they can provide lower limits and together with the good convergence behavior observed in the pressure and baryon-number density, this indicates that a critical point is unlikely at temperatures $T>135~$MeV for $\mu_B\leq 2 T$.
In the future it will be important to reduce the errors on the 6th-order coefficients and to determine higher order coefficients
for a reliable extension of these studies to larger chemical potentials and to improve on the estimates of the radius of convergence.

\section{Lines of constant physics and freeze-out}
\label{sec_loc}
Thermal conditions at the time of chemical freeze-out may be characterized by lines in the $T-\mu_B$ plane on which certain thermodynamic observables stay constant \cite{Cleymans:1999st,Cleymans:2005xv}.
Using the Taylor series for bulk thermodynamic observables from lattice QCD such lines of constant physics 
can be determined, e.g. as a parameterization of a ''line of constant $f$'' \cite{Bazavov:2017dus},
\begin{eqnarray}
T_f(\mu_B) = T_0 \left(1-\kappa_2^f 
\left( \frac{\mu_B}{T_0}\right)^2- \kappa_4^f \left( \frac{\mu_B}{T_0}
\right)^4\right) \; .
\label{Tf}
\end{eqnarray}
Results for lines of constant pressure, energy density and entropy density are shown in Fig.~\ref{fig_lcp}
for three initial sets of values fixed at $\mu_B=0$. The initial temperatures, $T = $ 145, 155 and 165 MeV, correspond to constant energy densities of $\epsilon = $ 0.203(27), 0.346(41) and 0.556(57) GeV/fm$^3$, indicating that the physics is substantially different on these three lines.
The lines for the three observables agree quite well and they also agree with current estimates on the curvature of the crossover line (black lines) \cite{Kaczmarek:2011zz,Endrodi:2011gv,Bonati:2015bha,Bellwied:2015rza,Cea:2015cya}.
Also shown are results of freeze-out \cite{Adamczyk:2017iwn,Floris:2014pta} parameters and hadronization \cite{Becattini:2016xct} temperatures extracted from particle yields measured in heavy ion experiments. These were obtained by comparing data with model calculations based on the hadron resonance gas model.

\section{Cumulant ratios of net baryon-number}
\label{sec_skew}
\begin{figure*}[thbp]
        \centering
        \includegraphics[height=0.34\textwidth]{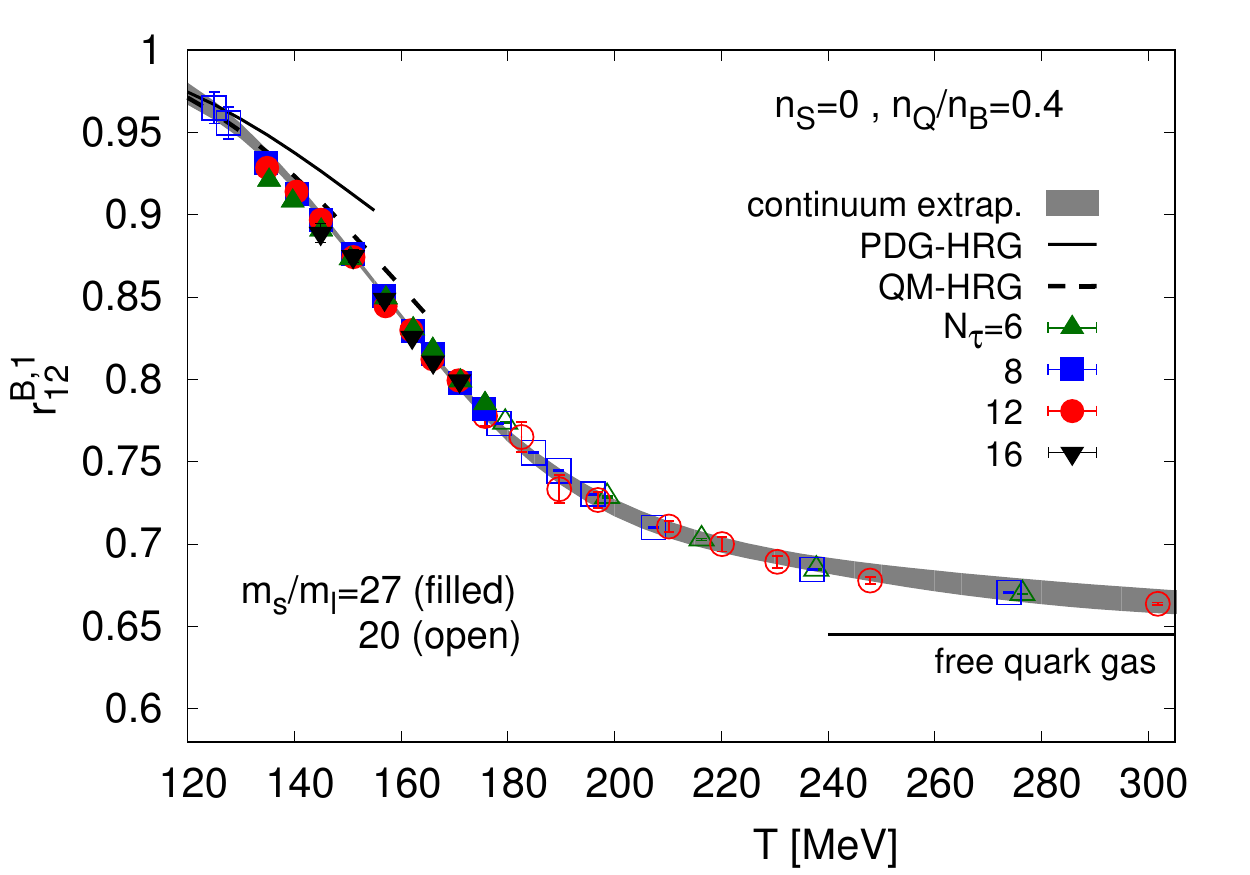}
        \includegraphics[height=0.34\textwidth]{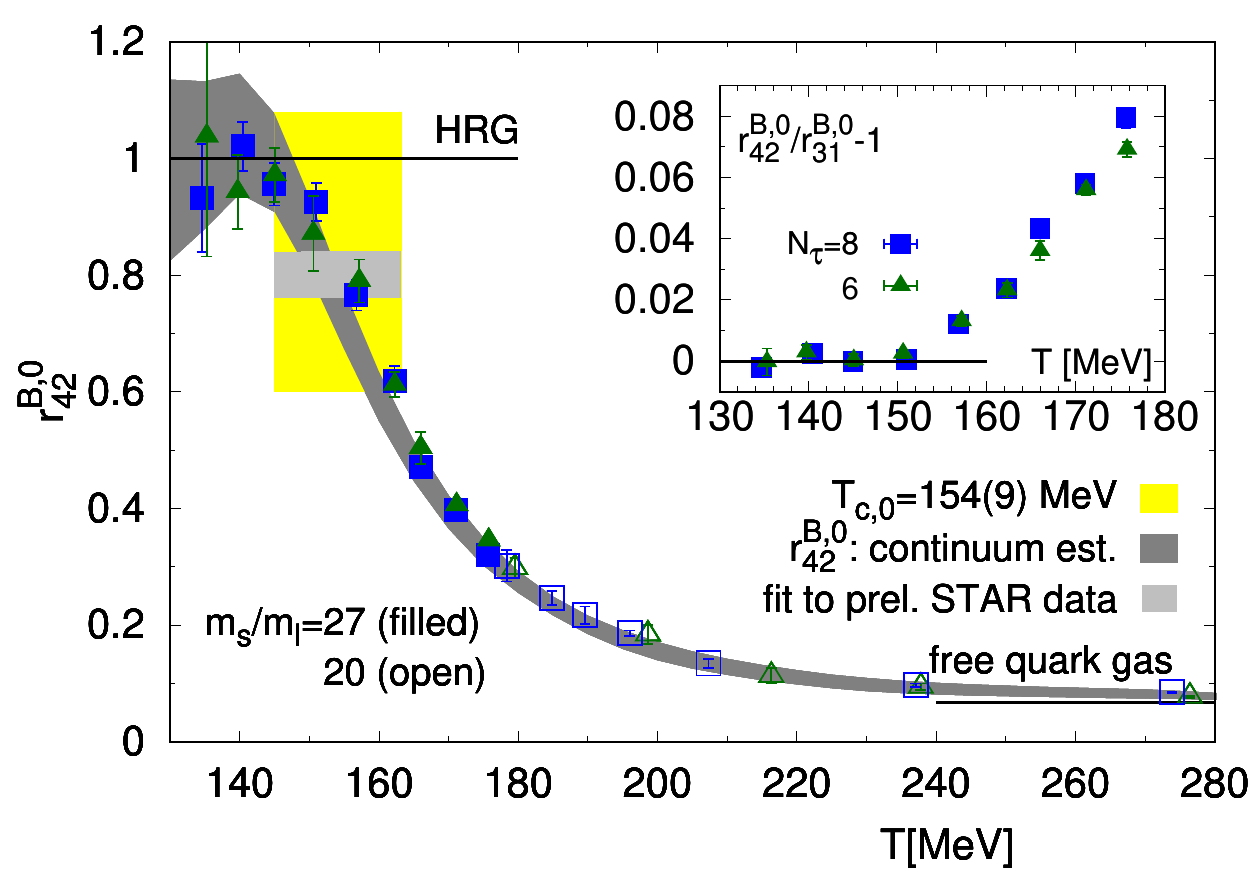}
        \caption{
Leading order expansion coefficients of the cumulant ratios $R_{12}^B$ (left)
and $R_{42}^B$ (right) calculated for strangeness neutral systems with an
electric charge to baryon-number ratio $r=0.4$ \cite{BNLBI_SKEW}.
%Right: Skewness and kurtosis ratios for net proton-number distributions versus
%mean net proton-number normalized by variance obtained by the STAR
%Collaboration at the BES \cite{Luo:2015ewa}.
All figures taken from \cite{BNLBI_SKEW}.
}
        \label{fig_skew}
\end{figure*}
The search for a possible critical point in the QCD phase diagram is one of the central goals of the beam energy scan at RHIC. Fluctuations and correlations among conserved charges play an important role in these studies
\cite{Hatta:2003wn,Ejiri:2005wq,Koch:2005vg,Adamczyk:2014fia}.
Net proton-number \cite{Aggarwal:2010wy,Luo:2015ewa,Thader:2016gpa} do not yet provide a clear evidence for the existence of a critical point, but the net proton-number fluctuations show a beam energy dependence which is so far not well understood. Using lattice QCD results on the temperature and baryon chemical potential dependence of net baryon-number fluctuations in equilibrium QCD thermodynamics may provide at least a qualitative understanding of these results.\\
Using $n^{th}$ order cumulants, $\chi_n^B(T,\mu_B)$, defined as partial derivatives of the QCD pressure with respect to the baryon chemical potential,
\begin{eqnarray}
\chi_n^B(T,\,u_B) = \frac{\partial^n P/T^4}{\partial(\mu_B/T)^n}\; ,
\end{eqnarray}
ratios of cumulants of net baryon-number fluctuations, i.e. mean ($M_B$), variance ($\sigma^2_B$), skewness ($S_B$) and kurtosis ($\kappa_B$) can be determined,
\begin{eqnarray}
R_{12}^B(T,\mu_B) &=&
\frac{\chi_1^B(T,\mu_B)}{\chi_2^B(T,\mu_B)}
= \frac{M_B}{\sigma_B^2}\\
R_{31}^B(T,\mu_B) &=&
\frac{\chi_3^B(T,\mu_B)}{\chi_1^B(T,\mu_B)}
= \frac{S_B \sigma_B^3}{M_B}\\
R_{42}^B(T,\mu_B) &=&
\frac{\chi_4^B(T,\mu_B)}{\chi_2^B(T,\mu_B)}
= \kappa_B \sigma_B^2\; .
\end{eqnarray}
For small values of $\mu_B$ these ratios can be expanded in terms of $\mu_B$, e.g. up to NLO,
\begin{eqnarray}
R_{12}^B(T,\mu_B) &=& r_{12}^{B,1} (\mu_B/T) + r_{12}^{B,3} (\mu_B/T)^3\\
R_{21}^B(T,\mu_B) &=& r_{31}^{B,0} + r_{31}^{B,2} (\mu_B/T)^2\\
R_{42}^B(T,\mu_B) &=& r_{42}^{B,0} + r_{42}^{B,2} (\mu_B/T)^2 \; .
\end{eqnarray}
In Fig.~\ref{fig_skew} we show the leading order expansion coefficients for the ratio of mean value and variance $(M_B/\sigma_B)$ (left) and kurtosis times variance $(\kappa_B \sigma^2_B)$ (right) for a strangeness neutral system, $n_S=0$, with electric charge to baryon-number ratio $n_Q/n_B=0.4$ obtained from (2+1)-flavor lattice QCD using the HISQ action \cite{BNLBI_SKEW}.\\
\begin{figure*}[thbp]
  \centering
        \includegraphics[height=0.32\textwidth]{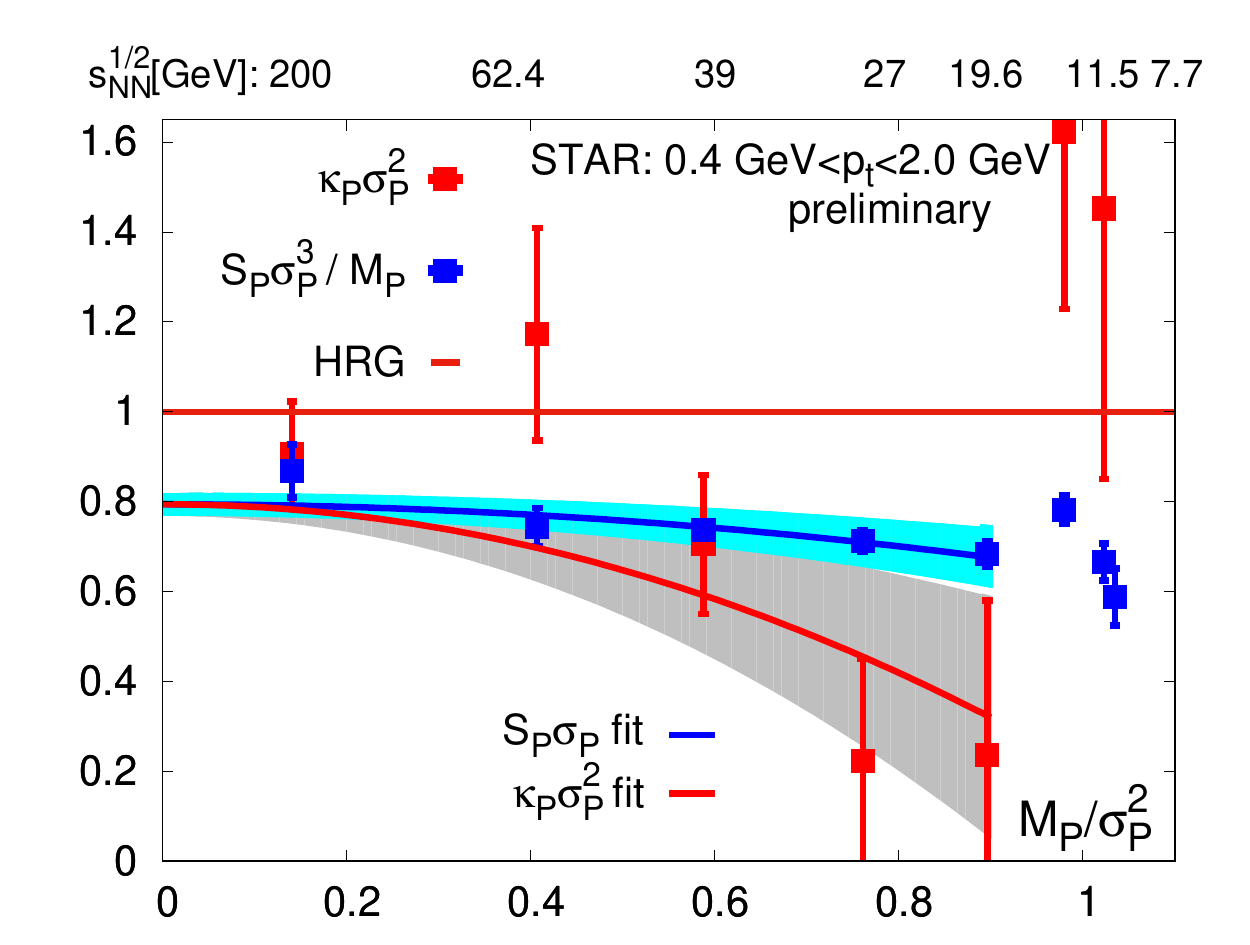}
        \hspace*{0.3cm}
        \includegraphics[height=0.285\textwidth]{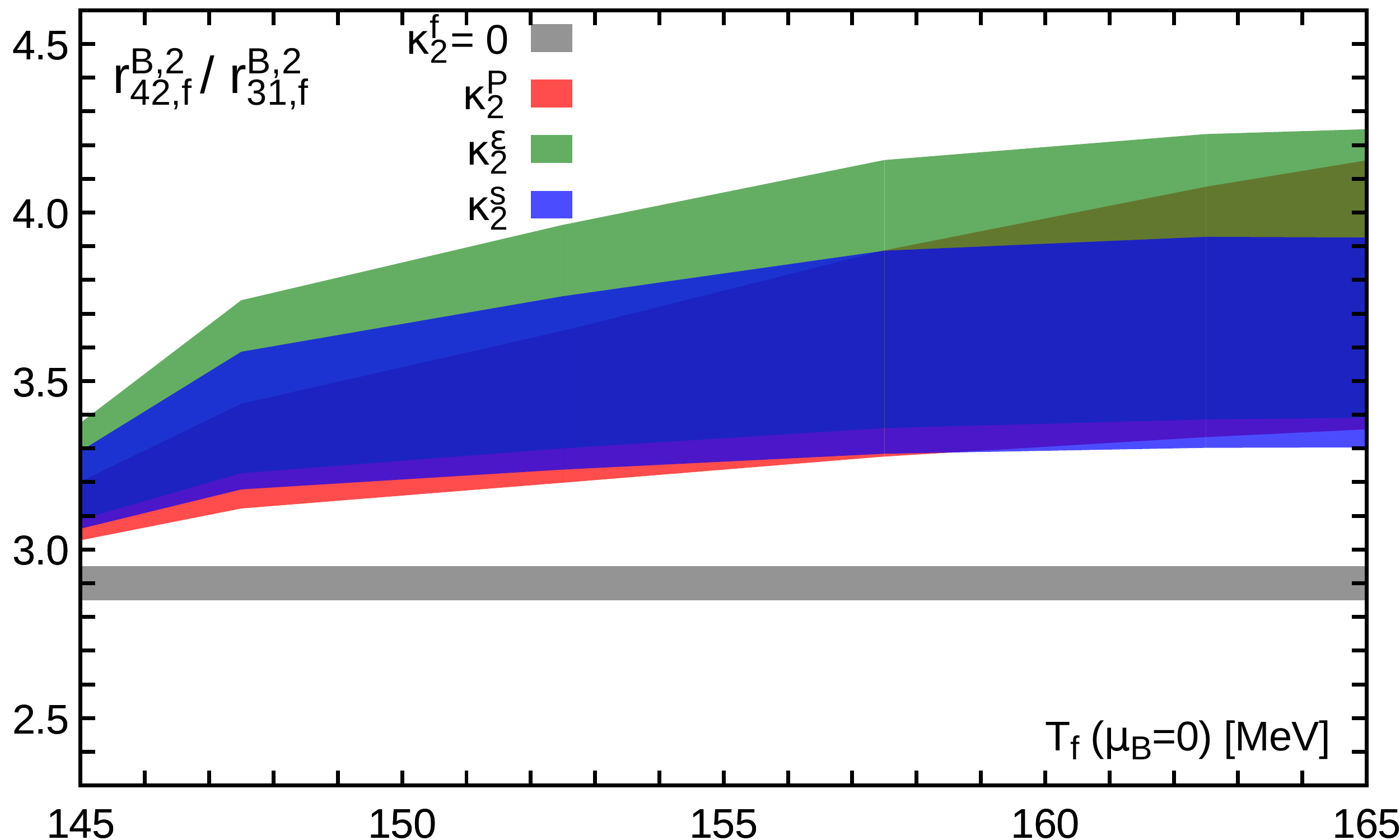}
        \caption{
Left: Skewness and kurtosis ratios for net proton-number distributions versus
mean net proton-number normalized by variance obtained by the STAR
Collaboration at the BES \cite{Luo:2015ewa}.
Right: Ratio of NLO coefficients of the skewness and kurtosis ratios on lines of constant pressure, energy density and entropy density.
All figures taken from \cite{BNLBI_SKEW}.
}
        \label{fig_skew2}
\end{figure*}
These results can be confronted with the preliminary data of the STAR Collaboration \cite{Luo:2015ewa} for the skewness and kurtosis ratios for net proton-number distributions shown in Fig.~\ref{fig_skew2} (left). Some qualitative features are apparent from this figure, which can not be described by a hadron resonance gas (HRG). ${S}_{P} \sigma^3_P/M_P$ is smaller than unity and shows a dependence on $\sqrt{s_{NN}}$. For small $M_P/\sigma_P^2$, i.e. large $\sqrt{s_{NN}}$ the relation ${S}_{P} \sigma^3_P/M_P\simeq \kappa_P\sigma^2_P$ seems to hold quite well, while with decreasing $\sqrt{s_{NN}}$, i.e. increasing $M_P/\sigma_P^2$, the kurtosis times variance $\kappa_B \sigma^2_B$ shows a stronger deviation from unity compared to the skewness ratio ${S}_{P} \sigma^3_P/M_P$. This behavior is clearly different from HRG model calculations, where both these ratios equal unity, while the leading order QCD results provide a natural explanation for these qualitative features.
The ratio of mean and variance, $M_B/\sigma_B^2$, in leading order is given by $r_{12}^{B,1} \mu_B/T$. Due to the positivity of the coefficient observed in Fig.~\ref{fig_skew} (left) it is a monotonically rising function of $\mu_B$, at least to leading order.\\
Already from the leading order coefficients the deviations of the skewness ratio $S_B\sigma_B^3/M_B$ from unity as well as the stronger decrease of the kurtosis $\kappa_B\sigma_B$ can be deduced.
From the inset of Fig~.\ref{fig_skew} we see that the difference of the
leading order expansion coefficients for skewness and kurtosis stays
small, which in the relevant temperature range for a comparison with
experimental data, 145~MeV $<$ T $<$ 165~MeV, 
leads to
\begin{eqnarray}
S_B\sigma^3_B/M_B \simeq \kappa\sigma_B^2
\end{eqnarray}
for small $\mu_B$, corresponding to large RHIC beam energies.\\
Assuming that the cumulant ratios for net proton-number distributions can qualitatively be compared to the QCD calculations of baryon-number cumulants, the intercept at $R_{12}^P=0$ in Fig.~\ref{fig_skew2}~(left) can be related to the QCD values in Fig.~\ref{fig_skew}~(right, gray band). This comparison leads to a freeze-out temperature of 153(3)~MeV which is in excellent agreement with the freeze-out temperature determined by the ALICE Collaboration from particle yields at the LHC \cite{Floris:2014pta} but differs significantly from the freeze-out parameters presented by STAR \cite{Adamczyk:2017iwn}.\\
In Fig.~\ref{fig_skew2}~(right) we show estimates for the ratios of slopes of the kurtosis to skewness calculated on lines of constant pressure, energy density and entropy density. These QCD predictions of a ratio of 3 - 4.5 in the relevant temperature regime compare well with an estimate of the ratio of slopes of 3.9(2.1) for the STAR data shown in Fig.~\ref{fig_skew2}~(left).\\
From this qualitative comparison of experimental data to QCD thermodynamics results, it is apparent that in the relevant parameter region for the freeze-out one should confront experimental data with QCD rather than with hadron resonance gas model calculations.

\section{Lattice constraints on the thermal photon rate}
\label{sec_photon}
\begin{figure*}[thbp]
        \centering
        \includegraphics[width=0.46\textwidth]{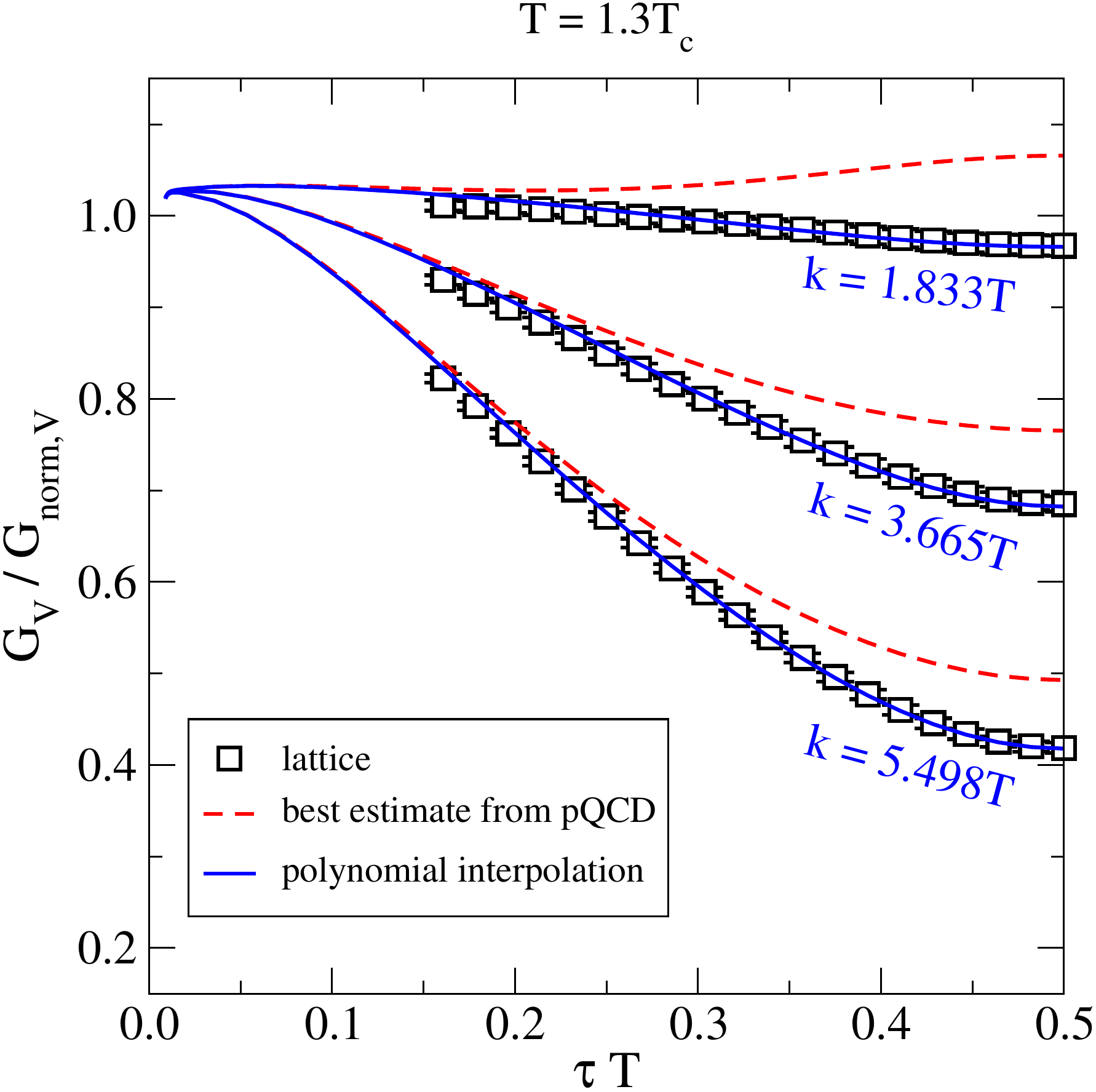}
        \hspace*{0.5cm}
        \includegraphics[width=0.46\textwidth]{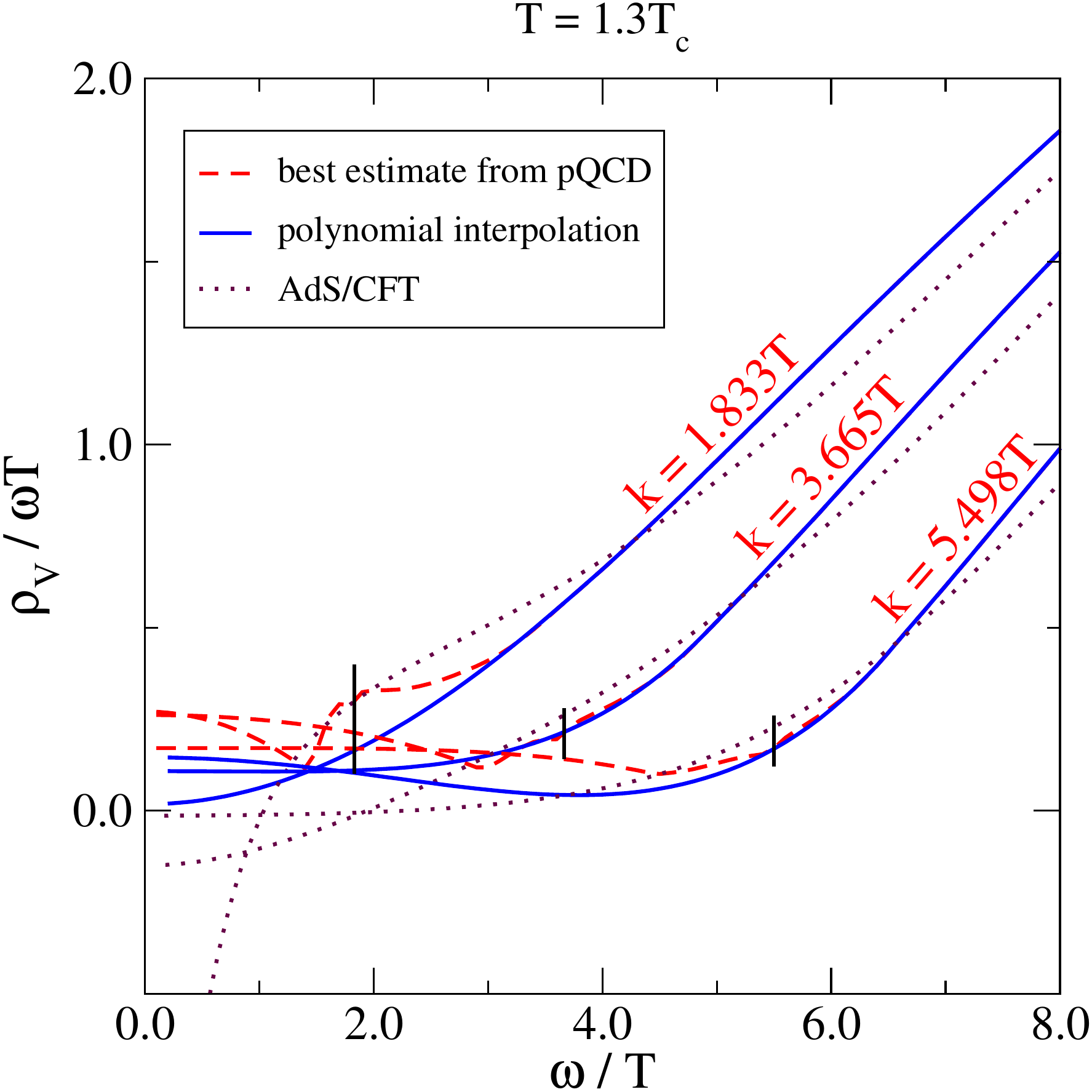}
        \caption{
          Left: Continuum extrapolated vector correlation function
          at non-zero momenta $k$ and fit result of the ''polynomial interpolations''. Also shown is the ''best estimate from pQCD'' based on \cite{Ghiglieri:2014kma,Ghisoiu:2014mha,Laine:2013vma}. Right:
          Corresponding spectral functions obtained from the fit. Vertical bars locate the light cone. Also shown in the AdS/CFT result from \cite{CaronHuot:2006te}, rescaled to agree with the QCD result at large $\omega/T$. All figures taken from \cite{Ghiglieri:2016tvj}.
        }
        \label{fig_photon1}
\end{figure*}
\begin{figure}[thbp]
        \centering
        \includegraphics[width=0.58\textwidth]{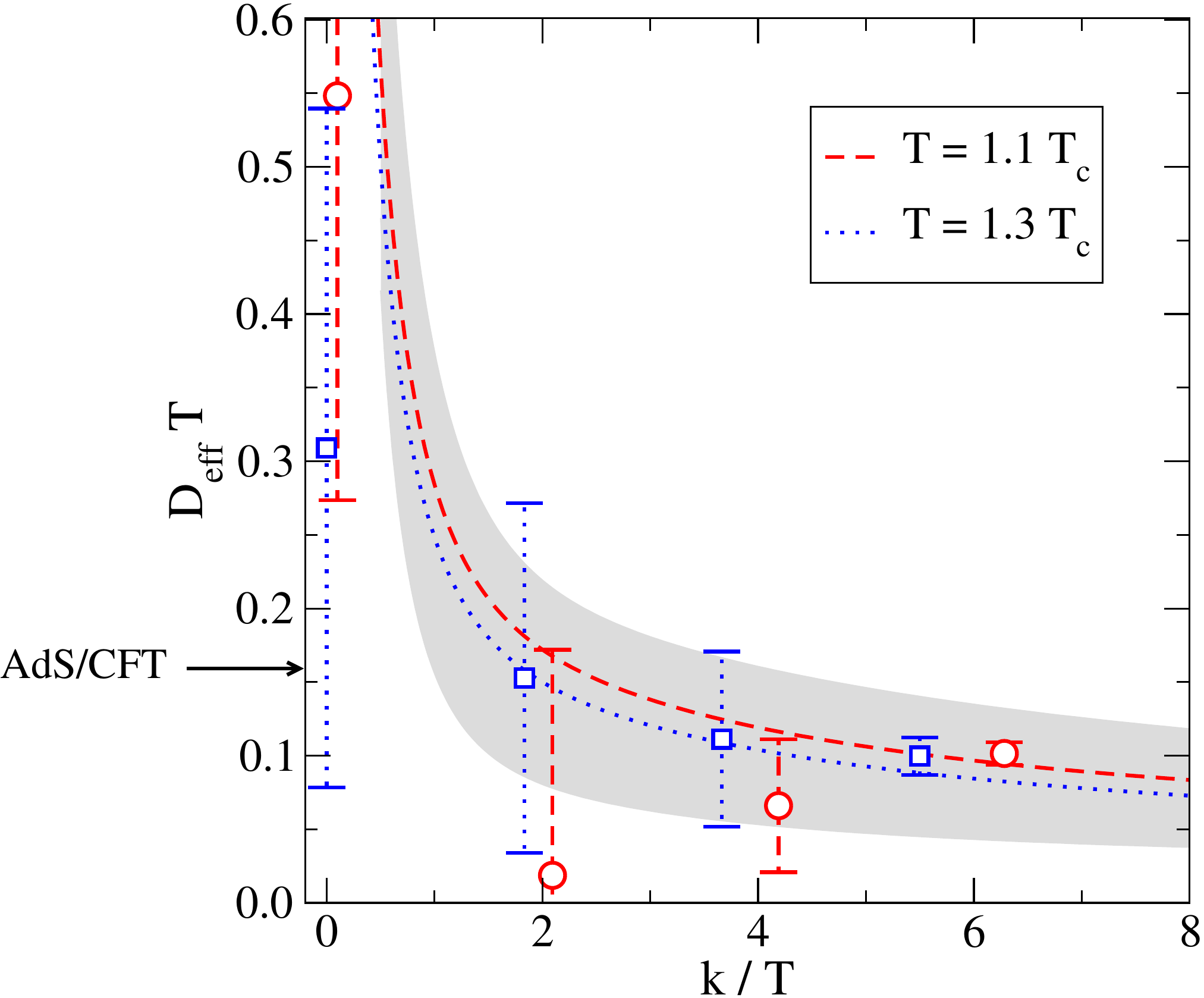}
        \caption{
          Lattice result for $D_{\mathrm{eff}}$, that determines the photon rate through (\ref{eq_photon2}), compared with the NLO perturbative prediction from \cite{Ghiglieri:2014kma} shown as a gray band.
          The AdS/CFT value is $DT=1/(2\pi)$ from \cite{Policastro:2002se}.
          Figure taken from \cite{Ghiglieri:2016tvj}.
        }
        \label{fig_photon2}
\end{figure}

Photons are excellent probes to study the properties of the strongly interaction matter at high temperature and density. Although the contribution of non-thermal photons are dominant for most invariant mass regions and one needs to integrate over all photon sources and the entire evolution of the medium, at intermediate invariant masses, $M$, thermal photons that are emitted from the QGP may provide important information on the interactions that the plasma particles experience.\\
Although a lot of progress has been made in weak-coupling perturbative calculations, in the interesting temperature regime probed in current heavy-ion collision experiments, QCD is still strongly coupled and non-perturbative methods required to obtain quantitative predictions. Nevertheless perturbative predictions play an important role to constrain the required spectral reconstruction of lattice QCD correlation functions.\\
Using large and fine lattices, currently only accessible in the quenched approximation, we have determined continuum extrapolated imaginary-time vector meson correlation functions at non-vanishing momenta \cite{Ghiglieri:2016tvj}. Some examples are shown in Fig.~\ref{fig_photon1}~(left) together with the currently best knowledge from perturbation theory. For short temporal distances, which are dominated by high frequencies in the spectral function, the lattice correlation functions approach these perturbative estimates and become even more perturbative at larger momenta.
Nevertheless non-perturbative effects become dominant at larger distances corresponding to small frequencies in the spectral function. We have used a polynomial interpolation for the spectral function that incorporates the expected linear behavior at frequencies much smaller than T and the best available perturbative predictions
\cite{Ghiglieri:2014kma,Ghisoiu:2014mha,Laine:2013vma},
up to ${\cal O}(g^2)$ for $M \gsim \pi T$ and up to ${\cal O}(g^8)$ for $M\gg\pi T$, and fitted the non-perturbative contribution to the lattice data.
The results give a good representation of the data as seen in Fig.~\ref{fig_photon1}~(left). The obtained spectral functions are shown in Fig.~\ref{fig_photon1}~(right). The values at the photon point are shown in Fig.~\ref{fig_photon2} in terms of $D_\rmi{eff}$ defined as
\begin{eqnarray}
 D_\rmi{eff}(k)  \;\equiv\;
 \left\{
 \begin{array}{ll}
    \displaystyle \frac{\rho^{ }_\rmii{V}(k,\vec{k})}{2 \chi^{ }_\rmi{q} k} 
 & \;, \quad k > 0 \\
    \displaystyle 
    \lim_{\omega\to 0^+}
 \frac{\rho_{ }^{ii}(\omega,\vec{0})}{3 \chi^{ }_\rmi{q} \omega}
 & \;, \quad k = 0 
 \end{array}
 \right.
 \;, \label{Deff}
\end{eqnarray}
which directly relates to the thermal photon rate,
\begin{eqnarray}
 \frac{{\rm d}\Gamma_\gamma(\vec{k})}{{\rm d}^3\vec{k}}
  \; = \;
 \frac{2  \alpha_\rmi{em} \chi^{ }_\rmi{q} }{3 \pi^2}
 \, \nB{}(k) D_\rmi{eff}(k)
 \;. \label{eq_photon2}
\end{eqnarray}
\label{sec:photon}
For momenta $k\gsim 3 T$ the results become comparable to the NLO perturbative prediction \cite{Ghiglieri:2014kma},
supporting the program to implement pQCD results into hydrodydamic codes \cite{Vujanovic:2013jpa,Burnier:2015rka,Paquet:2015lta},
though the results indicate the importance of non-perturbative effects at smaller $k$.
These results show the importance of combining perturbative results with continuum extrapolated correlation functions to obtain reliable estimates for spectral as well as transport properties. Although smaller momenta would be needed, the use of non-zero momenta may offer an alternative way to determine the diffusion coefficient as well as other transport coefficients.
It will be important for future studies
to fill the gap at small momenta and to study the effects of dynamical quarks especially for temperatures closer to $T_c$ where these will become important.

\section{Conclusions}
We have presented recent studies that extend the knowledge on the equation of state to non-zero baryon-number density, providing insights into the physics around freeze-out and as well as important input for hydrodynamic modeling of the evolution of the matter produced in heavy ion collisions.
A good convergence behavior is observed for the Taylor expansion of the pressure and baryon-number density for $\mu_B\leq 2 T$. Together with estimates on the radius of convergence of the Taylor series for net baryon-number fluctuations, this indicates that a critical point is unlikely at temperatures $T>135~$MeV for $\mu_B\leq 2 T$.
Qualitative features of the skewness and kurtosis ratios of net proton-number fluctuations measured by the STAR Collaboration can be understood from QCD results for cumulants of conserved baryon-number fluctuations.
In the future it will be important to improve on the calculation of higher order coefficients of the Taylor series
to extend these studies to larger chemical potentials and to improve the limits on a possible critical point in the QCD phase diagram.\\
As one example where progress has recently been made in the extraction of spectral and transport properties of the QGP,
constraints on the thermal photon rates from the QGP were discussed, based on a spectral reconstruction of continuum extrapolated lattice correlation functions constrained by recent perturbative results in the UV. These results become comparable to NLO perturbative predictions at $k\gsim 3 T$, but indicate non-perturbative contributions at small invariant masses.
It will be important to extend these studies to lower temperatures, even though the inclusion of dynamical quarks will become important and at some point possible vector meson resonances need to be included in the analyis.
Using continuum extrapolated lattice QCD correlation functions at non-zero momenta and incorporating perturbative information in the spectral reconstruction may provide an alternative way to extract transport coefficients in the future.

\section*{Acknowledgements}
The author would like to thank the members of the Bielefeld-BNL-CCNU Collaboration as well as Jacopo Ghiglieri and Mikko Laine for the fruitful collaborations. OK was partially supported by the DAAD. 

%\bibliographystyle{elsarticle-num}
%\bibliography{qm17}






% \end{thebibliography}

\end{document}